\documentclass[doublecol]{epl2} % for 2 columns style with line numbers
\usepackage{amsmath,amssymb}

\title{Observation of interspecies ion separation in inertial-confinement-fusion implosions}
\author{S. C. Hsu, \inst{1} T. R. Joshi,\inst{1} P. Hakel,\inst{1} E. L. Vold,\inst{1} M. J. Schmitt,\inst{1}
N. M. Hoffman,\inst{1} R. M. Rauenzahn,\inst{1} G. Kagan,\inst{1} X.-Z. Tang,\inst{1} 
R. C. Mancini,\inst{2} Y. Kim,\inst{1} and H. W. Herrmann\inst{1}}
\shortauthor{S. C. Hsu, T. R. Joshi, P. Hakel \etal}

\institute{                    
  \inst{1} Los Alamos National Laboratory, Los Alamos, New Mexico 87545, USA\\
  \inst{2} Physics Department, University of Nevada, Reno, Nevada 89557, USA}
  
\abstract{We report direct experimental evidence of
interspecies ion separation in direct-drive, inertial-confinement-fusion experiments on the OMEGA
laser facility.  These experiments, which used plastic capsules
with D$_2$/Ar gas fill (1\% Ar by atom), were designed specifically to reveal interspecies ion separation 
by exploiting the predicted, strong ion thermo-diffusion between ion species of large mass and
charge difference.  Via detailed analyses of imaging x-ray-spectroscopy data,
we extract Ar-atom-fraction radial profiles at different times, and observe both enhancement 
and depletion compared to the initial 1\%-Ar gas fill.
The experimental results are interpreted with radiation-hydrodynamic simulations
that include recently implemented, first-principles models of interspecies ion diffusion.
The experimentally inferred Ar-atom-fraction profiles agree reasonably, but not exactly, with calculated profiles
associated with the incoming and rebounding first shock.}

\begin{document}

\maketitle

%\section{Section title }

%paragraph 1
Interspecies ion separation has been proposed as a potential fusion-yield degradation mechanism
in inertial-confinement-fusion (ICF) implosions \cite{Amendt10prl,Amendt11pop,Bellei13pop}.  Several ICF experimental 
campaigns \cite{Rygg06pop,Wilson08jpcs,Dodd12pop,
Herrmann09pop,Casey12prl,Rinderknecht14pop,Rosenberg15pop,Rinderknecht15prl} have 
reported yield or yield-ratio anomalies where interspecies ion separation within the hot spot is a possible explanation.  
Meanwhile, first-principles analytic theories for multi-ion-species diffusion 
\cite{Amendt10prl,Amendt11pop,Amendt12prl,Kagan12prl,Kagan12pop,Kagan14pla,Molvig14pop} have been developed,
some of which have been implemented into ICF implosion
codes \cite{Gittings08csd,Vold15pop}, enabling us to quantitatively assess the magnitude
of interspecies ion separation and yield degradation in ICF implosions.  In addition, ion-Fokker-Planck
\cite{Larroche12pop,Inglebert14epl} and particle-in-cell kinetic
simulations \cite{Bellei13pop,Bellei14pop,Kwan15dpp}
are being used to address this problem, and are particularly appropriate for more-kinetic
scenarios, e.g., exploding pushers, hot-spot formation in ignition capsules, hohlraum plasmas, or near steep gradients at
shock fronts and material interfaces.  
We do not address the question here of
whether interspecies ion separation substantially affects the yield in
ignition-class implosions on the National Ignition Facility \cite{Edwards13pop}, but point
out that this and related work will help establish a validated
capability to answer the question in a quantitative manner.

%paragraph 2
The purpose of this work is to complement and expand upon
prior ICF experimental campaigns that relied on yield or yield-ratio anomalies
as evidence for species 
separation. Here, based on direct-drive ICF implosions on the OMEGA laser
facility \cite{Boehly97oc}, we provide direct, spatially resolved
experimental evidence of interspecies ion separation via detailed
analyses of imaging x-ray-spectroscopy data, which are less sensitive to other
potential causes of yield degradation in an ICF implosion.  In this work:
(1)~we exploit recently developed analytic theory 
\cite{Kagan12pop,Kagan14pla} of interspecies ion diffusion to design an ICF implosion
that maximizes interspecies ion diffusion, such that it is observable via x-ray diagnostics;
(2)~rather than focusing on separation between fuel species, e.g., D/$^3$He or D/T, we use 
D and a trace amount of Ar, where both the choice of Ar itself and its pre-fill concentration 
are chosen to maximize the expected interspecies diffusion coefficient; and (3)~we aim for a more-collisional
implosion so that interpretations using the recently formulated interspecies-ion-diffusion theories are appropriate.  
The main results in this Letter
are the first direct identification of interspecies ion separation in an ICF implosion based on the analyses of 
spatially resolved 
x-ray-spectroscopy data, observation of spatially non-uniform Ar-concentration enhancement and depletion
(compared to the spatially uniform 1~atom\% Ar pre-fill) during
the implosion, and reasonable agreement
of the experimentally inferred Ar-atom-fraction profiles with radiation-hydrodynamic simulations that
model interspecies ion diffusion from first principles.

\begin{figure}
\onefigure{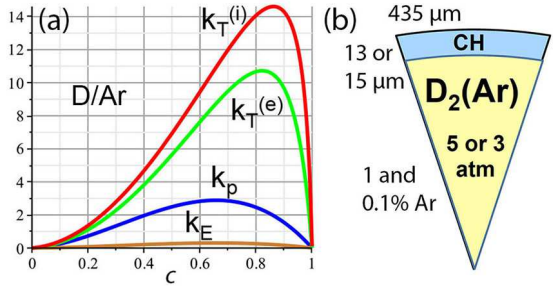}
\caption{(a)~D/Ar interspecies diffusion ratios vs.\ $c\equiv\rho_D/(\rho_D+\rho_{Ar})$ for
ion thermo-diffusion $k_T^{(i)}$, electron thermo-diffusion $k_T^{(e)}$, baro-diffusion $k_p$,
and electro-diffusion $k_E$, assuming charge state $Z_{Ar}=18$.
(b) We shot four target types:  15~$\mu$m/3~atm and 13~$\mu$m/5~atm, with 1 or 0.1~atom\% 
Ar.}
\label{fig:k_target}
\end{figure}

%paragraph 3
We briefly discuss the theory that guided these experiments.
The diffusive mass flux of the lighter ions relative to the center of mass in a plasma with two ion species
can be written as \cite{Kagan12pop,Kagan14pla}
$\vec{i} = -\rho D [ \nabla c + k_p \nabla \log p_i + (ek_E/T_i)\nabla \Phi + k_T^{(i)}\nabla \log T_i
+ k_T^{(e)} \nabla \log T_e]$,
%\begin{eqnarray}
%\mathbf{i} = && -\rho D \Big( \nabla c + k_p \nabla \log p_i + \frac{ek_E}{T_i}\nabla \Phi + k_T^{(i)}\nabla \log T_i\nonumber\\
%&&+ k_T^{(e)} \nabla \log T_e \Big),
%\label{eq:i}
%\end{eqnarray}
where $\rho$ is the total mass density, $D$ the classical diffusion coefficient,
$c\equiv \rho_l/\rho$ the lighter-ion mass concentration,
$p_i$ the ion pressure,
$T_i$ and $T_e$ the ion and electron temperatures, respectively, $\Phi$ the electric potential,
and $k_p$, $k_E$, $k_T^{(i)}$, $k_T^{(e)}$ the diffusion ratios for
baro-diffusion, electro-diffusion, ion thermo-diffusion, and electron thermo-diffusion, respectively;
$\vec{i}$ can also be written in terms of molar-concentration gradients~\cite{Molvig14pop}.
Because the theory predicts larger diffusion ratios for ion species with a large mass and charge
difference \cite{Kagan14pla},
we focused on a D/Ar mixture (rather than D/$^3$He or D/T)\@.
We chose Ar in particular because of its proven usefulness as an
x-ray spectroscopic tracer in ICF implosions \cite{Haynes96pre,Golovkin02prl,Regan02pop,Welser-Sherrill07pre,Nagayama11jap,Florido11pre,Nagayama14pop}.
We evaluated the diffusion ratios \cite{Kagan14pla}
for a D/Ar mixture versus $c$ [fig.~\ref{fig:k_target}(a)]; $k_T^{(i)}$ is the largest ratio
and maximizes at $c=\rho_D/(\rho_D + \rho_{Ar})\approx 0.84$, corresponding to Ar number
fraction $f_{Ar}\equiv n_{Ar}/(n_{Ar}+n_D) \approx (1-c)/(1+19c) \approx 1$\%.
Thus, we focused our attention on an implosion design
with large $\nabla T_i$ and $f_{Ar}\approx 1$\%
to maximize ion thermo-diffusion.  Furthermore,
we wanted relatively high $T_i$ and $D$ while keeping
the Ar--D mean free path much smaller than the hot-spot size for most of the implosion.  Although the theory
\cite{Kagan12pop,Kagan14pla,Molvig14pop}
is not strictly correct within a shock front, it should nevertheless capture the qualitative, leading-order effects
on interspecies ion separation during the passage and rebound of the first incoming shock.

\begin{table*}[tb]%[H] add [H] placement to break table across pages
\caption{Summary of shots analyzed for this Letter.  Yield (DD-n)
and burn-weighted ion temperature $\langle T_i \rangle$ are from the 12-m neutron time-of-flight scintillator, and 
neutron bang times (BT) 
and burn widths (BW) are from the cryogenic neutron temporal diagnostic (cryoNTD)\@.  Note that BT corresponds to
compression burn in these implosions.}
\label{table:shot_summary}
\begin{center}
\begin{tabular}{lllllllll}
Shot \# & laser & capsule OD ($\mu$m)/& D$_2$/Ar fill & Ar & yield & $\langle T_i \rangle$ & BT & BW \\
 & energy (kJ) & CH thickness ($\mu$m) & (atm) & (atom\%) & (DD-n) & (keV) & (ns) & (ps)\\
 78197 & 21.2 & 863.2/13.1 & 5.0 & 1.01 & 1.18E11 & 6.43 & 1.317 & 110\\
 78199 & 21.1 & 861.8/13.1 & 5.0 & 1.01 & 1.35E11 & 6.48 & 1.313 & 116\\
 78201 & 26.2 & 874.5/14.8 & 3.18 & 1.12 & 8.06E10 & 6.93 & 1.355 & 120\\
\end{tabular}
\end{center}
\end{table*}

%paragraph 4
We performed 1D HYDRA \cite{Marinak01pop} simulations of 
direct-drive OMEGA implosions with a 1-ns square pulse
to arrive at the capsule designs in fig.~\ref{fig:k_target}(b).  Table~\ref{table:shot_summary} summarizes
the as-shot laser and target parameters and neutron-based measurements.
To evaluate the amount of expected species separation, we also performed pre-shot 1D simulations using
xRAGE \cite{Gittings08csd} with its recently implemented two-ion-species transport model \cite{Molvig14pop}.
All our pre-shot xRAGE results showed relative deviations in $f_{Ar}$ of $\ge 20$\%, in some
cases much larger,
from the initial values ($f_{Ar}=0.01$ or 0.001) over much of the implosion, giving us confidence that we
could resolve the species separation
based on error bars achieved in previous analyses of imaging x-ray-spectroscopy data. 

%paragraph 5
The primary diagnostics are 
two x-ray multi-monochromatic imagers (MMI) \cite{Koch05rsi} with quasi-orthogonal views (mounted on
TIMs 3 and 4, where TIM stands for a
``ten-inch manipulator'' diagnostic port), a streaked 
x-ray spectrometer (TIM 1),
and a time- and space-integrated, absolutely calibrated x-ray spectrometer (TIM 2).
Standard neutron diagnostics and full-aperture backscatter systems were also fielded.  The MMIs used 10-$\mu$m-diameter pinholes, setting the spatial resolution at $\gtrsim 10$~$\mu$m, and recorded data
on x-ray framing cameras between 3.3--5.5~keV around the time of first-shock convergence.  Each camera recorded four frames per shot at different times; frame trigger times are given in the figure captions.
We obtained analyzable MMI data for several shots with initial $f_{Ar}\approx 1$\% 
(table~\ref{table:shot_summary}), and used the Ar-He$\beta$ (3.68~keV), %He$\gamma$ (3.87~keV), 
Ly$\beta$ (3.94~keV), and Ly$\gamma$ (4.15~keV) lines in our spectroscopic analyses.
Shots with initial $f_{Ar}\approx 0.1$\%, predicted to have much reduced ion thermo-diffusion,
provided weaker spectral lines.

\begin{figure}[htb]
\onefigure{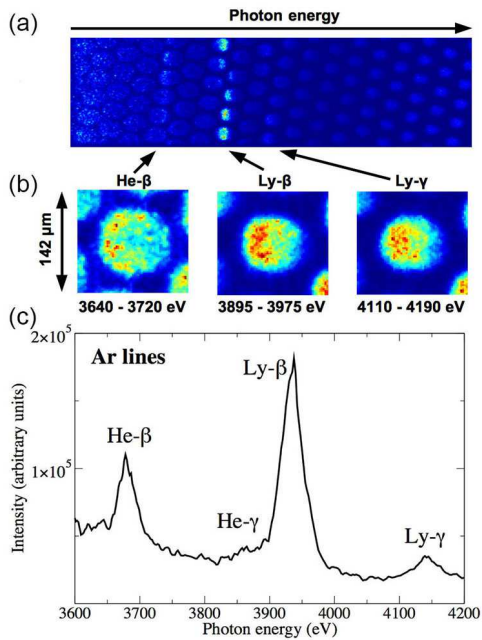}%
\caption{\label{fig2}(a)~MMI data showing an array of spectrally resolved
implosion-core pinhole images (intensity, arbitrary units) 
versus photon energy along the $x$ axis for shot 78199, TIM 3, frame 2 (1.14-ns trigger time),
(b)~Ar He$\beta$, Ly$\beta$, and Ly$\gamma$ narrow-band images constructed from 80-eV-wide (as
given below each image) vertical
strips of the data in (a). (c)~Space-integrated spectrum from (a) before
continuum subtraction.}
\end{figure}

\begin{figure}[htb]
\onefigure{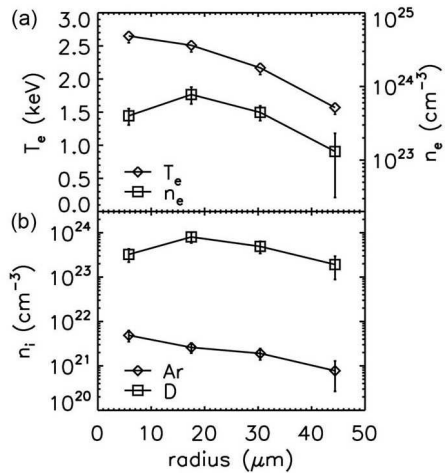}%
\caption{\label{fig3}(a)~Electron temperature and density and 
(b)~D and Ar ion densities versus radius, from analysis of MMI data
(shot 78199, TIM 3, frame 2, 1.14-ns trigger time).  See text for explanation
of error bars.}
%For these parameters, the Ar--D mean free path is of order
%the hot-spot size, assuming $T_i=T_e$.
\end{figure}

%paragraph 6
The key objective and result of this work is the experimental inference of $f_{Ar}$ versus radius $r$.  Deviation from
the spatially uniform target-pre-fill value constitutes proof of interspecies ion separation.
Figures~\ref{fig2} and \ref{fig3} illustrate the key steps of the MMI data processing and analyses required to arrive at 
$f_{Ar}(r)$.  MMI data processing to obtain narrow-band images [fig.~\ref{fig2}(b)] is based on the
method of refs.~\citenum{Nagayama11jap}, \citenum{Nagayama14pop}, and \citenum{Nagayama15rsi}.
Extraction of electron density $n_e$ and temperature $T_e$ versus $r$ are based on
the emissivity-analysis method of ref.~\citenum{Welser-Sherrill07pre}.
Figure~\ref{fig2}(a) shows one frame of MMI data after it has been converted from film density to 
intensity (arbitrary units).
%, using a 27-step wedge calibration.  
The photon-energy dependence of the MMI data is corrected using streaked
and absolutely calibrated spectral data \cite{flat-field}.
Figure~\ref{fig2}(b) shows the narrow-band images constructed from the
data in fig.~\ref{fig2}(a),
and fig.~\ref{fig2}(c) shows the space-integrated spectrum from fig.~\ref{fig2}(a) before continuum subtraction.

\begin{figure}[htb]
\onefigure{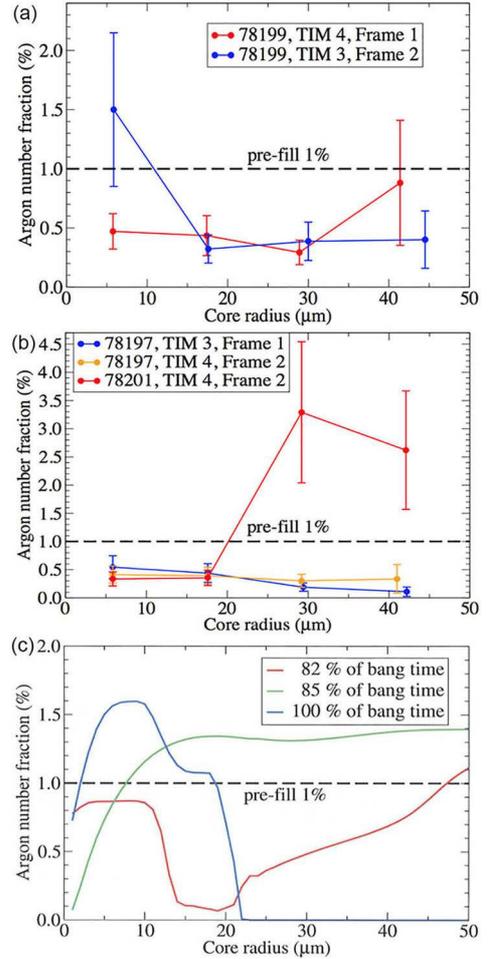}%
\caption{\label{fig4}Argon number fraction $f_{Ar}$ versus $r$ for (a)~shot 78199 (red $\approx 1.10$~ns = 84\%
of BT, blue $\approx 1.14$~ns = 87\% of BT), (b)~shots 78197 and 78201 (blue $\approx1.08$~ns = 82\% of BT, 
orange $\approx 1.09$~ns = 83\% of BT, red $\approx 1.18$~ns = 87\% of BT), and
(c)~1D post-shot
xRAGE simulation (shot 78199) including a two-ion-species transport model \cite{Molvig14pop}. }
\end{figure}

%paragraph 7
The next step is to extract $n_e$ and $T_e$ radial profiles.  Abel inversion is applied
to each narrow-band image, e.g., fig.~\ref{fig2}(b), giving argon
emissivity (with continuum subtracted) versus $r$ for four radial zones of approximately 10-$\mu$m width (set by
both the size of MMI pinholes and spectral signal-to-noise considerations for each zone),
at the time of the particular MMI frame.  The
time is approximately known based on the experimental trigger time of each frame of the x-ray framing camera, but due to
cable delays and the finite time of the sweep across the camera's photocathode strips, we regard the quoted times as having
an uncertainty $\sim 50$~ps.  The Ar-He$\beta$ and Ly$\beta$ emissivity profiles are analyzed to provide $n_e$ and $T_e$ versus $r$
[fig.~\ref{fig3}(a)] based on comparisons with detailed atomic-physics models.
The underlying atomic database was generated via the fully relativistic path
in the Los Alamos suite of codes RATS, ACE, and GIPPER \cite{Sampson09pr,Fontes15jpb}, followed by the collisional-radiative code 
ATOMIC \cite{Fontes15jpb,Abdallah90,Fontes16}.   
The resulting NLTE (non-local thermal equilibrium) spectral data were used in the intensity and emissivity analyses of the MMI data
[figs.~\ref{fig3}, \ref{fig4}(a), and \ref{fig4}(b)], 
as well as in the forward modeling of space-resolved spectra and narrow-band images
by the FESTR spectral post-processing code \cite{Hakel14pop,Hakel16cpc}.
Error bars for $T_e$ are obtained based on the greater of the $T_e$ resolution (100 eV) in the atomic database mentioned above 
or the standard deviation of $T_e$ (e.g., average of 63 eV for the four radial zones for shot 78199, TIM 3, frame 2) as determined from the theoretical emissivity ratio of Ly$\beta$/He$\beta$ at four different values of $n_e$
($5\times 10^{23}$, $8\times 10^{23}$, $1\times 10^{24}$, and $2\times 10^{24}$~cm$^{-3}$)
in our regime of interest.
Error bars for $n_e$ are obtained based on the greater of the uncertainties in the $T_e$ (100 eV) and
$n_e$ resolutions (1--$2\times 10^{23}$~cm$^{-3}$, depending on the density)
in the database, or the standard deviation in $n_e$ (e.g., average of $1.1\times 10^{23}$~cm$^{-3}$ for
the four radial zones for shot 78199, TIM 3, frame 2)
from assuming slightly higher and lower values of $n_e$ in the central zone 
(e.g., $4\pm 1 \times 10^{23}$~cm$^{-3}$ for shot 78199, TIM 3, frame 2), which
is derived from the line broadening of each of the three Ar lines from the space-integrated spectrum.

%paragraph 8
The argon density $n_{Ar}$ is determined using
$n_{Ar}=n_u/F_u(T_e,n_e)$, where $n_u$ and $F_u$ are the upper-level number densities and
fractional populations, respectively, of the particular line transition
being used \cite{Joshi15}.
$F_u(T_e,n_e)$ is retrieved from the atomic database based on the previously determined $n_e(r)$ and $T_e(r)$.
Since the spectral lines are sufficiently optically thin \cite{Welser-Sherrill07pre},
$n_u$ (arbitrary units) is extracted based on its proportionality to the corresponding 
observed line intensities, which are obtained from the space-resolved spectra extracted from 
annular regions of the implosion-core image by numerically integrating the area 
under the line after continuum subtraction.  Finally, 
the known pre-fill amount of Ar closes the system of linear equations relating $n_u(r)$ to $n_{Ar}(r)$
\cite{Joshi15}.   Error bars for $n_{Ar}$ are from
propagation of uncertainties in $n_u$ (e.g., 9\%, 6\%, 7\%, and 8\% from the innermost to outermost radial
zone for shot 78199, TIM 3, frame 2)
and $F_u(T_e,n_e)$ (e.g., 25\%, 22\%, 20\%, and 54\% from the innermost to outermost radial zone for
shot 78199, TIM 3, frame 2). The deuterium density $n_D$ is 
inferred from quasi-neutrality $n_e = Z_D n_D + Z_{Ar}n_{Ar}$, and finally $f_{Ar} \equiv n_{Ar}/(n_{Ar} + n_D)$.
Error bars for $n_D$ and $f_{Ar}$ use the propagated errors from $n_e$ and $n_{Ar}$, and $n_{Ar}$ and $n_D$,
respectively.
Figure~\ref{fig3} shows an example of $n_e$, $T_e$, $n_{Ar}$, and $n_D$ versus $r$, and figs.~\ref{fig4}(a) and \ref{fig4}(b) show
$f_{Ar}(r)$ from different shots and MMI views and times.  Results shown are the averages
from the Ar-He$\beta$, Ly$\beta$, and Ly$\gamma$ lines.  
The non-uniform spatial profiles of $f_{Ar}$ and their
deviation from the initial pre-fill value from several different shots and times
constitute proof of interspecies ion separation between the D and Ar, which is the main result of this
Letter.

%paragraph 9
To build confidence in our conclusions, we apply the same analysis procedure to synthetic
Ar-He$\beta$, Ly$\beta$, and Ly$\gamma$
images and space-resolved spectra.  The synthetic data are generated
using the FESTR code \cite{Hakel16cpc}, using the data in fig.~\ref{fig3} and assuming
1D spherical symmetry, for two cases:
(i)~$n_{Ar}$ as shown in fig.~\ref{fig3}(b) and (ii)~spatially uniform $f_{Ar}=0.01$ with equal $n_D + n_{Ar}$ as in (i).
In both cases, we extract $f_{Ar}(r)$ in good agreement with the $f_{Ar}(r)$ in
the synthetic data.  This demonstrates that our analysis 
accurately extracts $f_{Ar}(r)$ and is also
able to distinguish between species separation versus no species separation.

%paragraph 10
We also note some limitations of our analysis.  Firstly, our 
analysis assumes 1D spherical
symmetry, introducing possible errors if the implosion has significant 2D and/or 3D structure.
While our implosions appear fairly round, and quasi-orthogonal MMI views at similar times yield similar results
[e.g., shot 78197, blue and orange curves in fig.~\ref{fig4}(b)],
small-scale non-uniformities are clearly visible [e.g., fig.~\ref{fig2}(b)].  We assessed the effect of
non-uniformities for shot 78199 (TIM 3, frame 2)
by dividing the narrow-band images of fig.~\ref{fig2}(b) into three wedges of $120^\circ$ each and doing 
independent analysis on each wedge to obtain $n_e$ and $T_e$ profiles; the standard deviations of $n_e(r)$
and $T_e(r)$ across the three wedges are similar to the error bars already considered (as shown in
fig.~\ref{fig3}) and thus would not substantially increase the final error bars shown in fig.~\ref{fig4}.
Future work should employ both 2D \cite{Nagayama14pop} and 3D \cite{Nagayama12pop} reconstruction techniques to
further assess both 2D and 3D effects on our analysis.  Secondly, unaccounted-for
CH shell mix into the hot spot affects the species balance in the quasi-neutrality condition used in our analysis by 
causing overestimates of the inferred $n_D$, which affects the inferred $f_{Ar}$.
However, we have estimated that the local CH fraction in the hot spot
must exceed an unrealistically large value $\sim 10$\% before it 
could affect our conclusion of D/Ar separation presented in figs.~\ref{fig4}(a) and \ref{fig4}(b).  Finally, our use of the known initial amount
of Ar in the fuel region to infer absolute $n_{Ar}$ from $n_u$ (arbitrary units) means that $n_{Ar}$ is overestimated if
Ar leaks into the shell.  The latter could be a possible explanation for why $f_{Ar}(r) < 1$\%
everywhere for shot 78197 [fig.~\ref{fig4}(b)], keeping in mind that $f_{Ar}(r)$ is extracted only from regions
with visible, analyzable Ar spectral emission.  Resolution of the shell/gas-mix issues requires further work, including
the use of $N$-species ion-diffusion models that can model both the D/Ar separation and gas/shell mix.

%paragraph 11
Next, we interpret our results via a post-shot, 1D simulation of shot 78199 
using xRAGE \cite{Gittings08csd} with a two-ion-species transport model
\cite{Molvig14pop}.
Figure~\ref{fig4}(c) shows $f_{Ar}(r)$ for shot 78199,
revealing that $f_{Ar}$ is reduced ahead of the incoming first shock 
and enhanced behind it [green and red curves of fig.~\ref{fig4}(c)].  After shock reflection (at
$\approx 1.18$~ns and $\approx 87$\% of neutron bang time in the simulation), $f_{Ar}$ is enhanced
throughout much of the hot spot (up to $\approx 20$~$\mu$m) through neutron bang time [blue curve of fig.~\ref{fig4}(c)].  Furthermore, 
when ion thermo-diffusion, which is the strongest expected contributor to interspecies separation between D and Ar, is turned off
in the calculation, the $f_{Ar}$ enhancement and depletion largely disappear.  
Independently, these effects are also
seen in post-shot simulations from another radiation-hydrodynamics
code, to which a multi-ion-species diffusion model
has been added \cite{Hoffman15pop} and
recently updated to include ion thermo-diffusion
\cite{Schunk77rg,Zhdanov02,Paquette86apjss,Kagan14pla}.  Additional detailed 
comparisons between our data and simulation results from both codes will be reported elsewhere.
While detailed validation is beyond the scope of this work, note
that there is reasonable agreement between the data
and two key aspects of the post-shot simulation of shot 78199:  (1)~the 
peak magnitude of  $f_{Ar}$ ($\approx 1.5$\%), i.e., compare maxima of blue
curves in figs.~\ref{fig4}(a) and
\ref{fig4}(c), and (2)~the qualitative evolution of $f_{Ar}$, in that $f_{Ar}$ is below 1\% 
nearer the origin at an earlier time [red curves of figs.~\ref{fig4}(a) and \ref{fig4}(c)] and above 1\%
nearer the origin at a later time [blue curves of figs.~\ref{fig4}(a) and \ref{fig4}(c)].  Differences
in the details of the $f_{Ar}$-profile evolution between the observations and xRAGE simulation could arise
because (1)~there could be more D enhancement ahead of the shock front than the xRAGE 
prediction due to kinetic effects near the shock front, and (2)~shell mix, which affects $f_{Ar}$ through
an altered species balance in the quasi-neutrality condition, is not included in
the xRAGE simulation.

%paragraph 12
Additional evidence supports our identification of interspecies ion separation. The FESTR code
\cite{Hakel16cpc} is used to find the best spectral fit to radially resolved Ar spectral data.  In all cases examined,
when $f_{Ar}$ is allowed to be a free parameter in the fitting instead of being
fixed at the initial pre-fill value of 1\%, the fit is improved, e.g., 36\% improvement 
in normalized $\chi^2$ for the third radial zone (corresponding to $r\approx 30$~$\mu$m) of
the data in fig.~\ref{fig2}.

%paragraph 13
In conclusion, we have reported direct experimental evidence for interspecies ion separation in an ICF implosion,
via detailed analyses of imaging x-ray-spectroscopy data.  These direct-drive OMEGA
implosions of plastic capsules with D$_2$/Ar gas fill
were designed to maximize interspecies ion thermo-diffusion between ion species of large mass
and charge difference.  This campaign benefitted
from pre- and post-shot radiation-hydrodynamic
simulations including a first-principles treatment of interspecies ion diffusion.  The simulations show that $f_{Ar}$
is enhanced and depleted, respectively, in front of and behind the first incoming shock, with $f_{Ar}$ enhancement persisting
at the center from first-shock convergence through neutron bang time.
Our data agree reasonably well with the calculated effects on $f_{Ar}(r)$ due to the incoming
and rebounding first shock,
but more detailed validation, which is beyond the scope of this work, is needed.
The experimental Ar K-shell spectral signatures are overtaken by
bremsstrahlung continuum by the time of neutron bang time (around the time of peak compression),
and thus we are unable to report whether interspecies separation is observed through
bang time.  Nevertheless, 
this first result using x-ray spectroscopy to detect interspecies ion separation encourages further 
work toward establishing a validated capability to address its role in yield degradation of ignition-scale ICF implosions. 

\acknowledgments
We acknowledge R. Aragonez, T. Archuleta, J. Cobble, J. Fooks, V. Glebov, M. Schoff, T. Sedillo, C. Sorce, R. Staerker, 
N. Whiting, B. Yaakobi, and the OMEGA operations team for their support in experimental planning, execution,
and providing processed x-ray and neutron data.  This work was supported by the LANL ICF 
and ASC (Advanced Simulation and Computing) Programs under U.S. Department of Energy contract no.\ DE-AC52-06NA25396.

\bibliographystyle{eplbib.bst}
%\bibliography{ms.bib}

\begin{thebibliography}{10}
\bibitem{flat-field}
The MMI data is corrected as follows. First, the streaked x-ray-spectroscopy
  data (SSCA) is corrected by matching the time-integrated SSCA data to the
  space- and time-integrated, absolutely calibrated x-ray-spectroscopy data
  (XRS2). Then the space-integrated MMI spectrum is corrected to match the
  corrected SSCA data at the time of the MMI data.
\end{thebibliography}

\end{document}